# Title: Distinguishing a Majorana zero mode using spin resolved measurements

**Authors:** Sangjun Jeon*, Yonglong Xie*, Jian Li*, Zhijun Wang, B. Andrei Bernevig, and Ali Yazdani‡

**Affiliations:** Joseph Henry Laboratories and Department of Physics, Princeton University, Princeton, NJ 08544, USA

*These authors contributed equally to the manuscript.
‡email: yazdani@princeton.edu

**Abstract:**

One-dimensional topological superconductors host Majorana zero modes (MZMs), the non-local property of which could be exploited for quantum computing applications. Spin-polarized scanning tunneling microscopy measurements show that MZMs realized in self-assembled Fe chains on the surface of Pb have a spin polarization that exceeds that due to the magnetism of these chains. This feature, captured by our model calculations, is a direct consequence of the nonlocality of the Hilbert space of MZMs emerging from a topological band structure. Our study establishes spin polarization measurements as a diagnostic tool to uniquely distinguish topological MZMs from trivial in-gap states of a superconductor.

**Main Text:**

Localized Majorana zero modes (MZMs) are non-Abelian fermions that emerge at the edges of one-dimensional topological superconductors and may be used as building blocks for future topological quantum computers (*1–3*). A variety of condensed matter systems can be used to engineer topological superconductivity and MZMs (*4–7*). To date, evidence for MZMs has come from experiments that have detected their zero-energy excitation signature in various spectroscopic measurements (*8–10*). However, despite the recent progress, the question as to whether MZM could be uniquely distinguished from a finely-tunned or accidental trivial zero energy edge state in these experiment is currently unresolved (*11–13*). The interpretation of the experimentally observed zero energy modes as MZMs in various platforms has relied so far on showing that the zero energy mode is detected when the parameters of the system makes it most likely to be in a topological superconducting phase. While there is intense interest in utilizing such modes to create topological qubits (see for example, Ref. (*14*)), arguably, no unique experimental signature of MZMs is yet to be reported. Here we describe how spin-polarized spectroscopic mesurements (*15–21*) can be used to distinguish MZM from trival localized quasi-particles by showing features in such experiments that are a direct consequence of the nonlocality of the Hilbert space of MZMs emerging from a topological band structure. The spin polarization not only provides a litmus test for the existence of MZM, but also is a unique property of these novel qausi-particles that can be utilized for long-range transfer of quantum information between spin qubits (*22, 23*).

Previous studies have provided strong evidence that combining ferromagnetism in chains of Fe atoms together with the large spin-orbit coupling (SOC) interaction of the Pb(110) surface on which they are self-assembled gives rise to a proximity induced topological superconducting phase. This phase exhibits strongly localized MZMs at both ends of these chains (*9, 24–26*). High-resolution STM studies have directly visualized MZMs, placed a stringent bound on their splitting ($< 45\ \mu eV$), shown evidence for the predicted equal electron-hole spectral weight using spectroscopy with superconducting tips, provided a detailed understanding of their spatial profile, as well as revealed the robustness of MZM signatures when the Fe chains are buried by the deposition of additional superconducting material on top (*27*). Since trivial zero energy localized states may still show some of the features attributed to MZMs, we turn our attention to spin polarized measurements of this system.

We have performed spin-polarized STM measurements (*28*) of self-assembled Fe chains using Fe-coated Cr tips, which while ferromagnetic at their apex, do not have strong enough magnetic fields to disrupt the superconductivity of the Pb(110) substrate (the measurement schematic is shown in Fig. 1A, see also Fig. S1 and supplementary section 1). For our studies, before making measurements in zero applied magnetic field, the tip's magnetization is first trained in situ with the application of a magnetic field that is either parallel ($P$) or antiparallel ($AP$) to the vector normal to the surface. We have found that the observation of the hard BCS gap in the Pb substrate can be used as sensitive probe of trapped flux in our superconducting magnet, which can be canceled before proceeding the spin-polarization measurement (see Fig. S2). The field training switches the orientation of the STM tip's magnetization relative to that of the Fe chains, which results in a change of the apparent height of the chains and a uniform shift of the topographic features, as shown in Fig. 1B-D (see also Fig. S5). The observation of two types of contrast in topographs of many different chains with the same spin-polarized tip confirms the ferromagnetic behavior of our chains (see also Fig. S3 and S4). The height difference in the topographs reflects the difference between the "up" and "down" spin polarized density of normal states ($\rho_N^\uparrow, \rho_N^\downarrow$) of the chains near the Fermi energy ($E_F$), the integral of which (between $E_F$ and voltage bias $eV_B$) is related to the tunneling current which is kept constant by the STM feedback loop. This compensation of $\delta\rho_N = \rho_N^\uparrow - \rho_N^\downarrow$ by the so-called STM set-point effect, for $eV_B$ close to $E_F$, is best reflected in the spectroscopic measurements ($G(V) = dI/dV|_{eV=E} \propto \rho(E)$) of the Fe chains as shown in measurements in Fig. 1E, where a small 100 mT (out-of-plane) magnetic field is used to suppress superconductivity in Pb and Fe chains. The spectra measured at all locations on the chains near $E_F$ with both tip orientations ($G_P(V), G_{AP}(V)$) are equal, since the average spin polarization is compensated by adjusting the tip height for the two spectra (see also Fig. S6-S8 and supplementary section 2-4). It is possible to avoid this compensation and in fact to artificially enhance the influence of spin polarization on $G(V)$, by choosing a setpoint bias that is further from $E_F$ (see also Fig. S1 and S3 and supplementary section 1). Previously, we used this approach to demonstrate that our Fe chains are ferromagnetically ordered (30 meV in Ref. *9*). However, for the purpose of the current study, we focus on a small bias window near $E_F$, where the normal state spin dependent contribution is compensated by the STM set-point effect. As shown in Fig. 1F, the STM measured spin polarization $P(E) = (G_P(E) - G_{AP}(E))/(G_P(E) + G_{AP}(E))$, is zero over an energy window of ±5 meV. Such measurements therefore allow us to detect whether low energy quasi-particle states show any spin polarization beyond that due to ferromagnetism of our atomic chains.

Our finding that the MZM shows a unique spin signature is demonstrated by measurements of the spin-polarized zero bias conductance $G_P(0), G_{AP}(0)$ maps of the atomic chains in zero magnetic field, as shown in Fig. 2. The "double-eye" spatial structure of MZM in such high-resolution maps at zero bias near the end of Fe chains has been previously measured with un-polarized tips (27). It is theoretically understood using model calculations that consider both the spectral weight of the MZM in the superconducting substrate, as well as the trajectory of the STM tip during the constant current conductance map measurements. The data in Fig. 2, however, show that the magnitude of the MZM signature in $G_P(0)$ and $G_{AP}(0)$ maps depends on the magnetic polarization of the STM tip. Moreover, the contrast can be reversed by reorienting the tip's magnetization or by examining different chains with different ferromagnetic orientation (see Fig. S9 & S10 and supplementary section 5). We also note that $G_P(0)$ or $G_{AP}(0)$ maps do not show any contrast away from the end of the chain, where the signal in the bulk of the chain is likely dominated by the background from higher energy quasi-particles in our system, due to thermal broadening in our experiments (performed at 1.4K). The spin contrast shown in Fig. 2, reproduced for many chains, demonstrates that the end zero mode in our system has a polarization beyond the normal state background due to magnetism of the chains.

The details of the spin dependence of the tunneling conductance as a function of energy and position throughout the chain can be found in Fig. 3, which in addition to the spin polarization of the edge-bounded MZM, also shows the polarization of the other in-gap states at higher energies. These so-called Shiba states induced by magnetic structures on superconductors (29, 30) are predicted to be spin polarized (30, 31); however, their spin properties have not been experimentally detected. To better characterize the spin polarization of both the MZM and the Shiba states in spectroscopic measurements, in Fig. 4A & 4B we show averaged spectra at the end and in the middle of the chain along with the spin polarization $P(E)$ computed from these spectra. In addition to the MZM spin polarization signature as a peak at $P(0)$, another key finding is the antisymmetric behavior of $P(E)$ for Shiba states at higher energies. These measurements demonstrate that the states with energy $|E| < \Delta_{\mathrm{Pb}}$ , where $\Delta_{\mathrm{Pb}}$ is a superconducting gap, show features beyond those expected simply from the ferromagnetism of the chain, the average polarization of which is compensated by the STM set point effect as described above.

The first step in understanding that our results of spin polarized measurements uniquely distinguish the presence of MZM in our system is to show that no trivial localized state can give rise to a signal in the STM measured $P(E)$ at zero energy. To illustrate this point, we consider a single magnetic impurity in a host superconductor (Fig. 5A), which gives rise to a localized Shiba state within the superconductor's gap ($\Delta_{Pb}$) at $eV = \pm E_0 < \Delta_{Pb}$ (Fig. 5B). We have recently computed the spin properties of these states using a model that considers a magnetic quantum impurity (for example a partially filled $d$ level) coupled to a host superconductor (see Ref. *32*, and supplementary section 6). The changes in the superconducting host's normal state $\rho_N^{\uparrow/\downarrow}$ induced by the magnetic impurity via the exchange interaction determine the spin polarization of the Shiba state inside the gap: $\rho_S^{\uparrow/\downarrow}(E) = \pi \rho_N^{\uparrow/\downarrow} \sqrt{\Delta_{Pb}^2 - E_0^2} \, \delta(E \mp E_0)$ (Fig. 5B). From this result, we see that the polarization of a trivial localized Shiba state tuned to zero energy ($E_0 = 0$) is no larger than that of the normal state polarization induced by the magnetic impurity. Moreover, it can be shown, by symmetry, that spin contrast in our constant current conductance measurements $\delta G(E) = G_P(E) - G_{AP}(E)$ for the Shiba state must be antisymmetric as function of energy $\delta G(E) = -\delta G(-E)$, thereby requiring no spin contrast in spin-polarized STM measurements for $E_0 = 0$ Shiba states (see Ref. *32* and supplementary section 6). Our observation of a spin contrast for the edge mode of our atomic chains therefore excludes the possibility that this zero-energy mode is due to a purely localized Shiba state at zero energy. Since a Shiba state due to magnetic impurities is the only known trivial mechanism able to produce a localized in-gap differential conductance peak that disappears in the absence of superconductivity, our observation of a peak in $P(E)$ at zero energy uniquely identifies the MZM in our chains. Furthermore, the antisymmetry in $P(E)$ described above for quasi-particle states at energies larger than zero detected both at the end and in the middle of the chain (Fig. 3 & 4) is in clear agreement with our calculation of $P(E)$ from Shiba states induced in Pb by our magnetic Fe chain.

To further show that our spin polarization measurements distinguish between trivial quasiparticle states and topological MZM, we compare our experimental results to model calculations of chains of magnetic impurities embedded in a superconductor (see supplementary section 7). A chain of magnetic atoms induces a band of Shiba states within the host gap, which produces pairs of peaks at positive and negative bias in a $G(V)$ simulation (Fig. 4C & 4D). The width of these peaks is related to the bandwidth of the Shiba states and their dispersion. The Shiba bands produce signatures in spin polarized measurements very similar to that of a localized Shiba state discussed above. Specifically, the energy antisymmetric feature we have discussed above in $P(E)$ as being the hallmark of localized Shiba states also describes the behavior of extended Shiba states detected both at the ends and in the middle of the chains as shown in Fig. 4C & 4D. These features are reproduced in our experimental results (Fig. 4A & 4B), and confirm our understanding of signatures in $P(E)$ arising from the extended Shiba states in our chains.

Our theoretical considerations of a hybrid chain-superconductor system also reveal that a larger polarization than that of the normal state background, detected at zero energy in our experiments (Fig. 4A), is a unique signature of the non-local nature of MZMs. Unlike trivial localized states, MZMs can only exist at the edge of a chain system with extended electronic states, the properties of which not only dictate the formation of MZMs but also determine the spin polarization of the MZMs as well as that of the normal state background. Specifically, MZMs appear at the edges of ferromagnetic chains that have a normal state band structure with an odd number of spin-split bands crossing $E_F$. The MZM in such chains emerges from only one of the spin-polarized bands at $E_F$, and in our case, is associated with the exchange-split $d$-bands of the ferromagnetic Fe chains, as schematically shown in Fig. 5C. The large band splitting (~2eV) between majority and minority bands in the Fe chains estimated in our previous study (*9, 24*) guarantees the minority spin bands crossing $E_F$. The spin polarization of the MZM (Fig. 5D) is directly related to the spin-polarization of these specific momentum states at $E_F$, where the induced topological gap is opened, although spin-orbit coupling results in a very slight tilting of the spin polarization away from the ferromagnetic magnetization. To understand why this localized MZM spin polarization can be detected in our STM measurements, we must contrast the MZM polarization to that of the normal state background, which is compensated by the STM set-point effect. The normal state background spin polarization of our system involves not only the Fe states crossing $E_F$, but also states further away in energy that are broadened due to the hybridization with the Pb substrate (Fig. 5C & 5D). The sum contribution of these delocalized states throughout the chain to the background at its edge is always smaller than that of a localized MZM (*32*). Consequently, as our calculations for a hybrid chain-superconductor system show in Fig. 4C & 5D, the MZM spin polarization exceeds that of the background normal state providing an important test for the detection of MZMs in our system.

We contrast the nature of the background relevant to the MZM spin polarization measurements to that of a trivial localized state, which may accidently appear near the end at near zero energy. For such a localized state, as in the case of an isolated Shiba impurity described above, the relevant normal state background is local, resulting in compensation by the STM set-point effect and therefore showing no contrast in the spin-polarized measurements. A zero energy peak in measurements of $P(E)$ in spin-polarized STM (Fig. 4A) is due to non-local nature of MZM background and a unique signature of these novel excitations allowing us to distinguish them from trivial edge modes.

Looking beyond our atomic chain Majorana platform, spin-selective spectroscopy measurements using quantum dots have been recently proposed for the semiconducting nanowire Majorana platform (*33*). As we have studied here for the atomic chains, such experiments are expected to distinguish between trivial and non-trivial edge modes and probe the non-local nature of MZMs in the nanowire platform. The spin polarization of MZMs may also provide a useful approach to create highly polarized spin currents and the key to entangle these topological localized quantum states with conventional spin qubits. In fact, there are proposals outlining how a hybrid system of spin and MZM qubits can be used to perform universal quantum computation (*23*). The possibility that electron tunneling between spin qubits (based on quantum dots or individual defects) and MZMs can realize a quantum superposition between spin qubits and MZMs and vice versa is intriguing. Such a process can, for example, facilitate long-distance entanglement between spatially well-separated spin qubits (*22*). The first step in exploring the properties of a hybrid system of MZMs and spin qubits is the detection of the MZM spin polarization, as we have demonstrated here.

**Acknowledgments:**

We acknowledge discussions with L. Glazman, F. von Oppen, K. Franke, P. Lee, and C. Kane. This work has been supported by ONR-N00014-14-1-0330, ONR-N00014-11-1-0635, ONR- N00014-13-1-0661, Gordon and Betty Moore Foundation as part of EPiQS initiative (GBMF4530), NSF-MRSEC programs through the Princeton Center for Complex Materials DMR-142054, NSF-DMR-1608848, Simons Investigator Award, NSF EAGER Award NOA - AWD1004957, DOE-BES, Packard Foundation, ARO-MURI program W911NF-12-1-046, and Eric and Wendy Schmidt Transformative Technology Fund at Princeton. This project was also made possible using the facilities at Princeton Nanoscale Microscopy Laboratory. BAB wishes to thank Ecole Normale Superieure, UPMC Paris, and Donostia International Physics Center for their generous sabbatical hosting.

**Supplementary Materials:**

Materials and Methods

Supplementary Text: Section 1-7

Fig S1 – S10

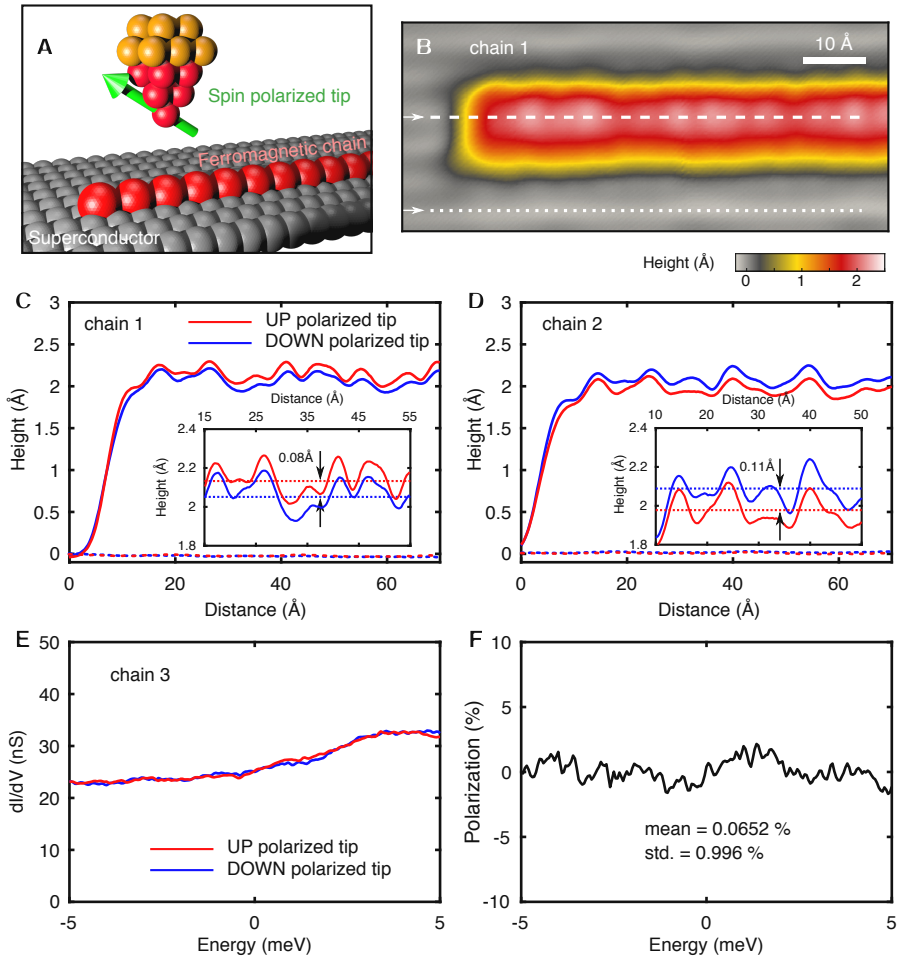

**Fig. 1. Spin polarization of Fe chains on Pb(110) at normal state.** (**A**) Schematic of the SPSTM measurement on a MZM platform probed by Fe/Cr tip whose magnetization can be controlled by external magnetic field. (**B**) Topography of a typical Fe chain (chain 1) under zero magnetic field ($V_{set}$=10 mV, $I_{set}$=750 pA). (**C** and **D**) Height profile of chain 1 (C) and chain 2 (D) with UP polarized tip (solid red curves) and DOWN polarized tip (solid blue curves). Dashed lines show the height profiles of the Pb substrate measured with both tips. The positions of the topographic profiles are shown in (B). Insets show a magnification of the height profile on chains. Red and blue dashed lines in the insets show the average height of the chain measured with UP and DOWN polarized tip, respectively. (**E**) Spectra measured in the middle of chain 3 at 100 mT with UP (red curve) and DOWN (blue curve) polarized tips which correspond to $G_P^N(V)$ and $G_{AP}^N(V)$ ($V_{set}$=-5 mV, $I_{set}$=500 pA, $V_{mod}$=40 µV). (**F**) Calculated polarization of the spectra shown in (E).

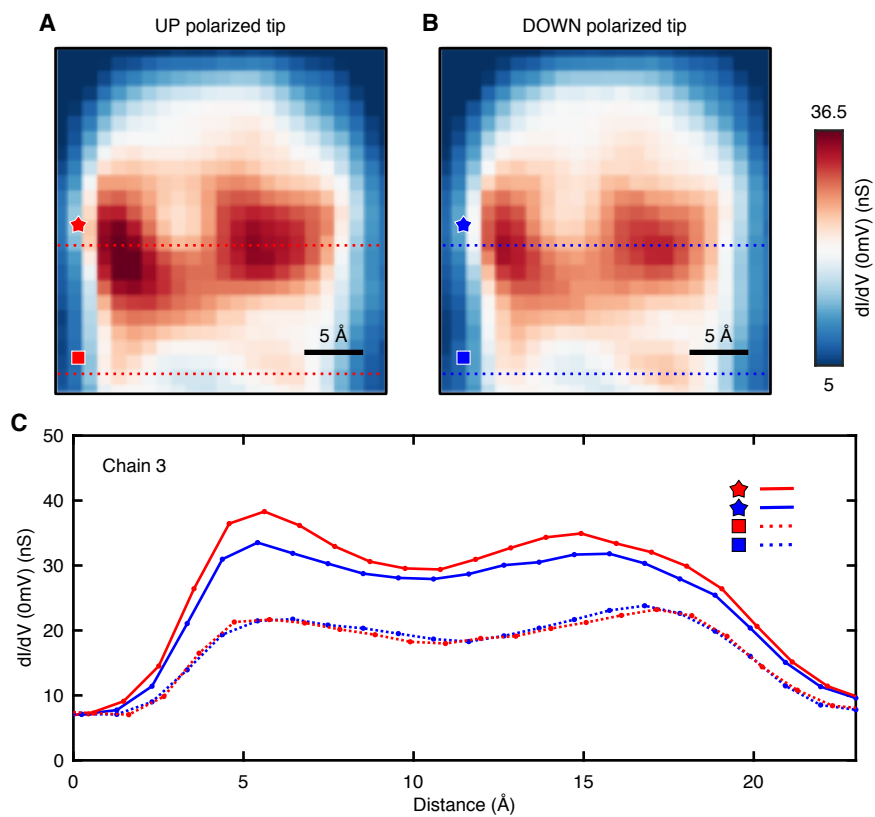

**Fig. 2. Spin-polarized zero energy end state.** (**A** and **B**) Zero energy conductance map near the chain end taken with UP (A) and DOWN (B) polarized tips ($V_{set}$=-5 mV, $I_{set}$=500 pA , $V_{mod}$=40 μV). The conductance at double-eye feature shows enhanced contrast between (A) and (B). (**C**) Line cuts taken from (A) and (B). Solid lines represent the conductance profile across the double-eye feature measured with UP (red curves) and DOWN (blue curves) polarized tips. Dashed lines show typical conductance profiles away from the chain end.

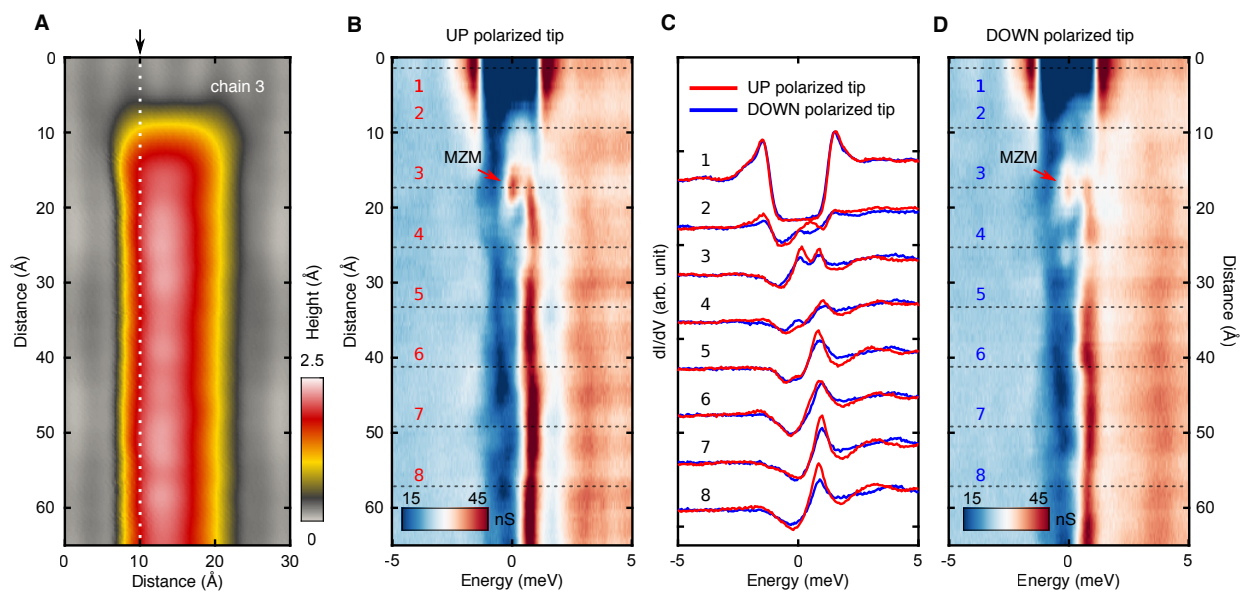

**Fig. 3. Spatially resolved spin-polarized states on a Fe atomic chain.** (**A**) Topographic image of chain 3 taken at zero external field ($V_{set}$=-10 mV, $I_{set}$=750 pA). (**B** and **D**) Spatial variation of the spectra taken along the white dashed line shown in (A) with UP (B) and DOWN (D) polarized tips ($V_{set}$=-5 mV, $I_{set}$=500 pA, $V_{mod}$=40 μV). (**C**) Individual spectra from (B) (red curves) and (D) (blue curves) at the labeled location. End state appears at spectrum 3.

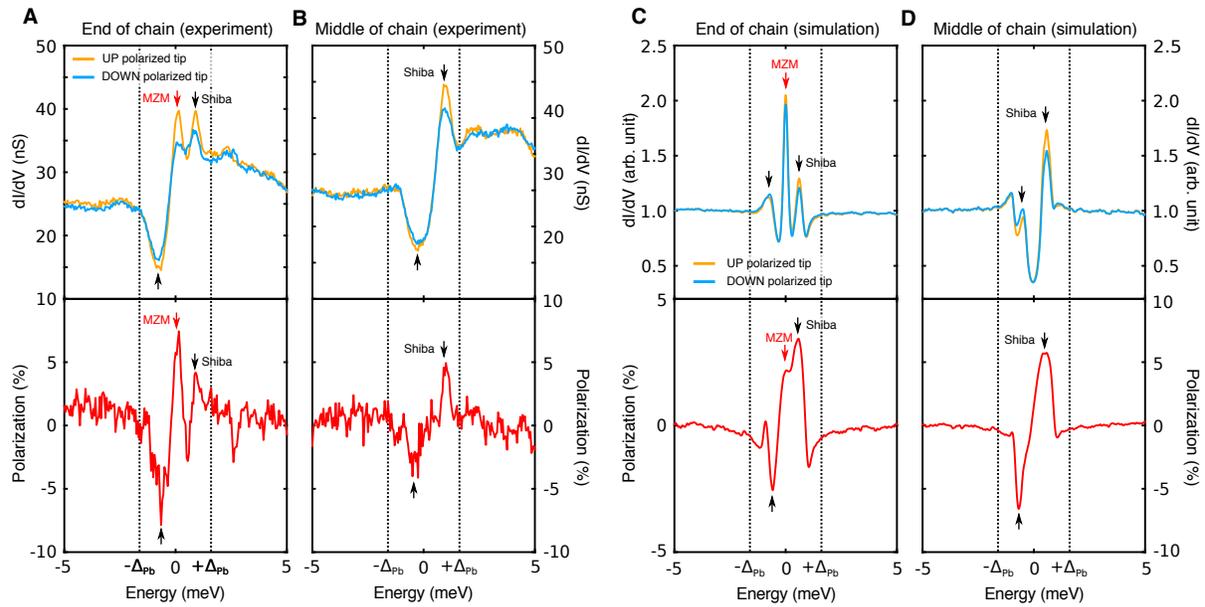

**Fig. 4. Comparison of experimental data and SP-STS simulation.** (**A** and **B**) experimentally obtained spectra at the end of chain (A) and in the middle of chain (B) and their corresponding polarization ($V_{set}$=-5 mV, $I_{set}$=500 pA, $V_{mod}$=40 μV). Yellow and blue curves are taken with UP and DOWN polarized tip, respectively. Red arrows mark the zero-energy end state and black arrows mark the van-hove singularity of the Shiba band. (**C** and **D**) Simulated spectra at the end of chain (C) and in the middle of chain (D) and their polarization. Both experimental and simulated results show strong positive polarization for the end state and energetically antisymmetric polarization for the Shiba states at higher energies.

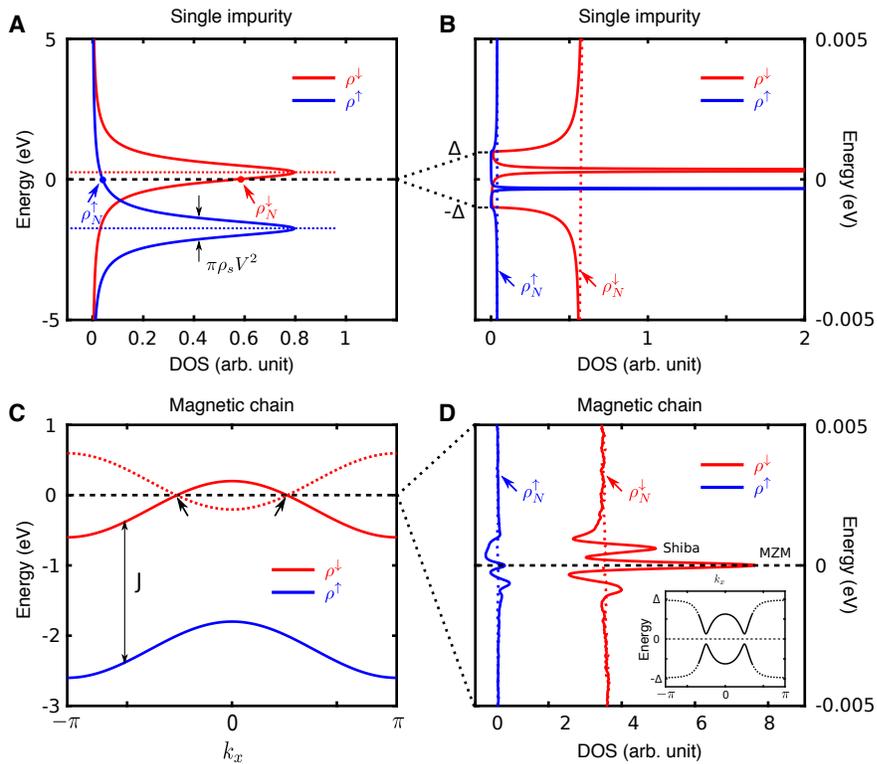

**Fig. 5. Single magnetic impurity and magnetic chain model calculation.** (**A** and **B**) Model calculation for a magnetic impurity hybridized with superconductor (see supplementary section 6 for the description of model Hamiltonian used in this calculation, $M = 1, \mu = 0.75, v = 0.4, \Delta = 0.001, \eta = 5e^{-6}$). The original $d$-orbital states of impurity are marked as dashed lines in (A) (red for minority state and blue for majority state). Solid curves represent the spin DOS of hybridized impurity-superconductor system. A pair of Shiba states inside a superconducting gap is shown in (B). Dashed lines in (B) correspond to the normal state spin DOS of minority state (red) and majority state (blue) which are the same value as spin DOS at the Fermi energy marked in (A). (**C** and **D**) Model calculation for a magnetic chain embedded in a superconductor (see supplementary section 7). Solid curves in (C) are the dispersion of original $d$-orbital bands (red for minority band and blue for majority band). Red dashed curve represents the hole-copy of the minority band. Black arrows mark the momenta where minority band crosses the Fermi level. Spin DOS calculated at the end of chain displayed in (D) show MZM and van-hove singularities of Shiba bands. Red and blue dashed lines are normal state spin DOS of minority and majority bands, respectively. Inset shows the dispersion of Shiba bands.

# Supplementary Materials for

Distinguishing a Majorana zero mode using spin resolved measurements

Sangjun Jeon, Yonglong Xie, Jian Li, Zhijun Wang, B. Andrei Bernevig,
and Ali Yazdani

correspondence to: yazdani@princeton.edu

**This PDF file includes:**

**Materials and Methods**

The STM measurements were performed on a homebuilt ultra-high vacuum scanning tunneling microscope operating at T=1.4 K, which can store two samples simultaneously. The Pb (110) single crystal surface was prepared by repeated cycles of argon ion sputtering (E = 1 keV for 15 minutes) and annealing (at T = 250 °C for 40 minutes). Atomic chains were synthesized by 0.1-0.5 monolayer deposition of Fe on the Pb surface at 100°C and light post-annealing (T = 175°C for 7 minutes).

We used bulk Cr tips that were conditioned in situ on Fe clusters to retain spin polarization at zero external magnetic field and then used to image Fe chains on Pb. The differential conductance maps are recorded simultaneously to STM topography using standard lock-in technique with a modulation voltage $V_{mod}$ (rms, f= 4 kHz) and a time constant of 5ms. dI/dV curves are recorded with open feedback using an appropriate bias $V_{set}$ and current $I_{set}$ to stabilize the tip-sample distance. We train our tip by applying B= ±1T to obtain two different tip magnetizations at zero magnetic field.

**Supplementary Text**

Section 1. Spin-polarized scanning tunneling microscopy (SP-STM) measurement

Our Fe-coated Cr tip was prepared by conditioning a bulk Cr tip in situ on Fe cluster. The tip magnetization shows ferromagnetic behavior as shown in Fig. S1A. It follows the direction of external magnetic field at high enough field strength ($H \gg H_c \sim 0.25\ T$). At an intermediate field strength ($H \sim H_c$), the orientation of tip magnetization is defined by the competition between the external magnetic field and the underlying magnetic anisotropy of bulk Cr. The zero-field magnetization is pinned by the easy axis of the underlying magnetic anisotropy of bulk Cr which, in general, has a component perpendicular to the sample surface. To get an 'UP' ('DOWN') polarized tip, we train the tip by applying 1 T (-1 T) perpendicular to the sample surface and tune to zero field afterward. We check the spin polarization of the tip by comparing the conductance measured with two polarized tips both on Pb substrate and Fe chain. Since Fe is magnetic and Pb is non-magnetic, the conductance measured with two polarized tips away from the Fermi level show contrast on Fe chain and no contrast on Pb as shown in Fig. S1D & S1G. Here we are using the set bias away from the Fermi level to enhance the effect of spin-polarization on the conductance measurements with the STM under constant current condition, as described in the main text. The conductance change on Pb substrate at higher field compared to that at zero field is attributed to the spin-orbit coupling of the Pb surface (*9*). For zero-field measurements, we check the superconducting gap on Pb substrate before proceeding the measurement. If a full superconducting gap is absent due to trapped flux in the magnet, we add a small field to the magnet (~5mT) to compensate the trapped flux (Fig. S2).

To further confirm that the switching event in the hysteresis (Fig. S1A) is ascribed to the tip magnetization, we have measured the conductance maps of five chains (chains 5 to 9) with the same spin polarized 'UP' and 'DOWN' tips shown in Fig. S3. The chain 5, 6 and 9 show higher conductance when measured with 'UP' polarized tip (trained with positive 1 T), whereas the chain 7 and 8 show higher conductance when measured with 'DOWN' polarized tip (trained with negative 1 T). These measurements strongly indicate that the magnetizations of Fe chains are spontaneously chosen after the growth and demonstrate our capability of controlling the tip magnetization. This is consistent with DFT calculations that estimate the coercive field of these chains as ~ 1.4 meV per Fe atom (9). To demonstrate the stability of our spin-polarized tip, we have carried out the time sequences of three measurements of two chains (chain 10 and 11 in Fig. S4). While chain10 and 11 are located 1000 angstroms away and the measurements of chain 10 were conducted weeks apart, the two spin polarized measurements of chain 10 show exactly the same contrast (Fig. S4 A&B and E&F).

Figure S5 shows the topographies of chain 1 and chain 2 measured with 'UP' and 'DOWN' polarized tips. The topographic profiles measured on Pb(110) do not show any detectable contrast (dotted curves in Fig. S5C and S5F are magnified by 10 times) as expected on a non-magnetic substrate in contrast to that on Fe chains.

Section 2. Normal state spin-polarization of Fe chain

We can quantitatively estimate the normal state spin polarization of the Fe chains from STM measurements and compare it with the theoretical calculations that consider both the structure of these chains as well as the hybridization of their electronic structure with the Pb substrate. In these measurements, we have to consider how the STM feedback loop adjusts the tip height to keep the total tunneling current constant. This results in the normalization of the conductance that needs to be taken account to obtain the spin polarization of the chains. As pointed out in the main text, the 'UP' and 'DOWN' tips have different tip-sample distance depending on their magnetic alignment (parallel or antiparallel) with the Fe chains. The normalization can only be reliably determined if the set bias ($V_{set}$) is chosen such that the energy window is small to have a uniform normal state spin DOS in energy, but high enough that the normalization is dominated not by the in-gap states but by the normal state spin DOS. Practically, we chose $V_{set}$=-5 meV in this study which shows the best cancellation between spectra measured with two polarized tips (Fig. S6A & S6B) and is still far away from the superconducting gap edge of the Pb substrate. To extract the absolute spin polarization from 'UP' and 'DOWN' spectra, we carry out dI/dZ measurements to get the tunneling barrier parameter $\kappa$ as described in the formula:

$$I_{set}(V) = A e^{-\kappa z} \int_0^{eV} d\varepsilon \left[ \rho_t^\uparrow(\varepsilon) \rho_s^\uparrow(\varepsilon - eV) + \rho_t^\downarrow(\varepsilon) \rho_s^\downarrow(\varepsilon - eV) \right] \quad (S1)$$

where $\rho_t^{\uparrow,\downarrow}$ ($\rho_s^{\uparrow,\downarrow}$) are the spin dependent LDOS of the tip (sample), $A$ is a proportional constant. Both polarized tips show the same $\kappa$ of -2.30 (1/Å) on Fe chain within measurement error (Fig. S6C). By combining the extracted $\kappa$ and the height difference of the chain, which is typically 0.1 (Å), measured with 'UP' and 'DOWN' polarized tips, we find the absolute spin polarization to be 11.5 % for our tip-sample junction. We also carry out the DFT calculation for the Fe chain on the Pb(110) surface with three Fe atoms in a unit cell (as shown in the inset of Fig. S6D). The calculations were performed on a 9-atomic-layers Pb slab with 10 Å vacuum space between slabs. The positions of Fe and the nearest Pb atoms are optimized. The spin-resolved DOS is calculated without considering spin-orbit coupling. Fig. S6D shows the calculated spin DOS of the Fe chain whose polarization is 34 % at the Fermi energy. This is the upper bound of the polarization, when the tip is fully polarized and its magnetization is perfectly aligned with that of the sample.

Section 3. Influence of tip height on STM spectra

To quantify the conductance change of the in-gap states influenced by the tip height difference from 'UP' and 'DOWN' polarized tip, we carry out spectroscopic measurements in the middle of a chain (Fig. S7A) and at the end of a chain (Fig. S7B) with various junction resistances. Figure S7C summarizes the conductance change as a function of junction resistance at $\Omega_+$, $\Omega_-$, and ZBP energies. In contrast to the Shiba states, the normalized ZBP conductance is enhanced as the tip moves closer to the chain by changing the set current. This can be understood by the fact that the MZM is highly localized in contrast to the spatially extended Shiba states. Moreover, in the case of antiparallel (parallel) tip-sample configuration, the feedback loop would enhance(weaken) the conductance of MZM, whereas our measurements in this configuration show lower (higher) conductance for MZM. Decreasing the tip height by 1.0 (Å) will increase the tunnel current of 26 % which will give rise to a conductance enhancement of less than 1 % for the ZBP. Hence, the conductance difference measured by the 'UP' and 'DOWN' polarized tips is originated from the spin polarization of states in Fe chain.

Section 4. Spatially resolved spin-polarization of a Fe chain at 100 mT

As shown in Fig. 1E, the spin contrast in the normal state spectra is compensated by the STM feedback loop. Here we illustrate this cancelation of the spectra all along the chain (chain 3) at 100 mT (out-of-plane) – a field which suppresses the superconductivity (Fig. S8). The measurement is carried out on the same chain displayed in the Fig. 2 and 3 with the same atomic tip.

Section 5. Spin-polarized measurement on a Fe chain with opposite ferromagnetic orientation

Figure S9 shows the spin polarized measurement on chain 4 which has the opposite magnetization of chain 3, so 'DOWN' polarized tip shows stronger conductance than 'UP' polarized tip for the ZBP as well as the Shiba states located at positive energies. Since the spin polarization of MZM directly originates from the minority $d$-orbital band crossing the Fermi level, Fe chains with opposite ferromagnetic orientation hold MZM with opposite spin polarization accordingly. More examples of point spectra measured with different tips on different chains can be found in Fig. S10.

<u>Section 6.</u> <u>Single</u> <u>magnetic</u> <u>impurity</u> <u>model</u>

We have recently examined the spin properties of the in gap states induced by a single chain of magnetic atoms embedded in a superconductor (*32*). Here we highlight some of the important findings of these calculations. In particular, we explain the antisymmetric behavior of the spin polarization for an individual magnetic impurity. This leads us to conclude that if the in-gap Shiba state from such an impurity locates at zero energy, such a state would show no contrast in the measurement of P(E) as described in the main text. To illustrate these points, we consider the following BdG Hamiltonian of the hybrid system:

$$\hat{H} = \hat{H}_s + \hat{H}_d + \hat{H}_T \, , \tag{S2}$$

$$\hat{H}_s = \int d\boldsymbol{k} \left( \boldsymbol{c}_{\boldsymbol{k}}^\dagger \ \overline{\boldsymbol{c}}_{\boldsymbol{k}}^\dagger \right) H_s(\boldsymbol{k}) \begin{pmatrix} \boldsymbol{c}_{\boldsymbol{k}} \\ \overline{\boldsymbol{c}}_{\boldsymbol{k}} \end{pmatrix}, \tag{S3}$$

$$\hat{H}_d = \left( \boldsymbol{d}^\dagger \ \overline{\boldsymbol{d}}^\dagger \right) H_d \begin{pmatrix} \boldsymbol{d} \\ \overline{\boldsymbol{d}} \end{pmatrix}, \tag{S4}$$

$$\hat{H}_T = \int d\boldsymbol{r} \left( \boldsymbol{c}_{\boldsymbol{r}}^\dagger \ \overline{\boldsymbol{c}}_{\boldsymbol{r}}^\dagger \right) V \, \delta \, (\boldsymbol{r}) \tau_z \begin{pmatrix} \boldsymbol{d} \\ \overline{\boldsymbol{d}} \end{pmatrix} + h.c. \, , \tag{S5}$$

where

$$H_s(\boldsymbol{k}) = (t_s k^2 - \mu_s)\tau_z + \Delta\tau_x \, , \tag{S6}$$

$$H_d = M\sigma_z \otimes \tau_0 - \mu\sigma_0 \otimes \tau_z \, , \tag{S7}$$

$M$, $\mu$, $\Delta$, and $V$ represent the exchange energy, the chemical potential, the pairing potential, and the coupling strength between the impurity and the superconductor, respectively. $\boldsymbol{c}^\dagger$ and $\overline{\boldsymbol{c}}^\dagger$ represent the Nambu particle and hole creation operators for the superconductor, and $\boldsymbol{d}^\dagger$ and $\overline{\boldsymbol{d}}^\dagger$ represent the creation operators for one impurity. The $\sigma_z$ operators act in the spin space. The model adopted in the current paper is a "quantum Shiba" model whereas the Eq. S10 of ref (*9*) is a "classical Shiba" model which faithfully reproduces the quantum states on the Fe site. The superconductor Hilbert space, created by the operators c – in both S10 of ref (*9*) and in the current manuscript - is coupled to the d-orbital state of the Fe spin through the term $\hat{H}_T$ . The d-orbital of the Fe has a spin-splitting of order M, due to the Fe magnetization. While this Fe spin splitting will induce a splitting in the Pb sites which sit in close proximity to the Fe impurity in a "quantum Shiba" model, we introduce such a splitting by hand in a "classical Shiba" model where the d-orbital states of Fe are not present, but where the spin-splitting induced by the Fe on the Pb atoms close to the Fe impurity is introduced as a Zeeman-like term of magnitude J (= 2M).

The retarded Green's function relevant to the *d*-orbital is given by

$$G_d(E^+) = \left[ E^+ - H_d + \nu \frac{E^+ \tau_0 + \Delta \tau_x}{\sqrt{\Delta^2 - (E^+)^2}} \right]^{-1} , \tag{S8}$$

$$\nu = \pi \rho_s V^2 , \tag{S9}$$

where $E^+ = E + i\eta$ with $\eta$ a positive infinitesimal, and $\rho_s$ is the normal DOS of the superconductor at the Fermi energy. The third term in the square braket of the above equation is the self-energy coming from the coupling between the impurity and the superconductor. The spin density of the $d$-orbital is defined as

$$\rho^{\uparrow/\downarrow}(E) = \text{Tr}\left[ p_{\uparrow/\downarrow} A_d(E) \right] , \tag{S10}$$

where $p_{\uparrow/\downarrow} = \frac{1}{2}(\sigma_0 \pm \sigma_z) \otimes \frac{1}{2}(\tau_0 + \tau_z)$ and $A_d(E) = \lim_{\eta \to 0} \frac{i}{2\pi}[G_d(E^+) - G_d(E^-)]$. In the case of the low energy ($|E| < \Delta$) regime, the spin density can be simplified to

$$\rho^{\uparrow/\downarrow}(E) \simeq \pi \rho_N^{\uparrow/\downarrow} \sqrt{\Delta^2 - E_0^2} \; \delta(E \mp E_0), \tag{S11}$$

where

$$\rho_N^{\uparrow/\downarrow} = \frac{\nu/\pi}{(\mu \mp M)^2 + \nu^2} , \tag{S12}$$

$$E_0 \simeq \Delta \frac{M^2 - \mu^2 - \nu^2}{\sqrt{(M^2 - \mu^2 - \nu^2)^2 + 4M^2\nu^2}} , \tag{S13}$$

with the assumption of $M, \nu \gg \Delta$. The original spin up and down levels of the $d$-orbital located at $\mp M - \mu$ hybridize with the surrounding superconductor to give rise to a finite spin DOS at Fermi level – defined as $\rho_N^{\uparrow/\downarrow}$ (see Fig. 5A). The spin DOS of the in-gap states are dominated by the Shiba states located at $\pm E_0$ (Fig. 5B) and their weight is proportional to the normal state spin DOS ($\rho_N^{\uparrow/\downarrow}$) due to the total weight conservation. Since $\delta(E)$ is a symmetric function in energy, we find the normalized spin DOS of the in-gap states, defined as

$$\tilde{\rho}^{\uparrow/\downarrow}(E) = \frac{\rho^{\uparrow/\downarrow}(E)}{\rho_N^{\uparrow/\downarrow}}, \tag{S14}$$

satisfies the condition of $\tilde{\rho}^{\uparrow}(E) = \tilde{\rho}^{\downarrow}(-E)$, and hence $\tilde{\rho}^{\uparrow}(0) = \tilde{\rho}^{\downarrow}(0)$.

In a spin-polarized STM measurement, the tunneling current obtained by two opposite orientation spin polarized tips, denoted by $N$ and $P$, can be written as

$$I_{N/P}(x, z, V) = \sum_{\sigma = \uparrow, \downarrow} w_{N/P, \sigma}(z) \int_0^{eV} dE \, \rho^{\sigma}(x, E) , \tag{S15}$$

where $w_{N/P,\uparrow/\downarrow}$ are weight factors describing the tunnel ratios between the $N/P$ polarized tips to the spin $\uparrow/\downarrow$ DOS of the sample which include the information associated with the relative angle between the spin polarization of the tips ($N/P$) and the chain ($\uparrow/\downarrow$). In a STM measurement, the tip height is adjusted to keep the same total current, hence the measured spectra are normalized by the set current $I_{N/P}(V_{set})$ as follows

$$G_{N/P}(x,E) = \frac{\widetilde{w}_{N/P}\rho^{\uparrow}(x,E) + \rho^{\downarrow}(x,E)}{\widetilde{w}_{N/P}R^{\uparrow}(x) + R^{\downarrow}(x)} \quad , \qquad (S16)$$

where

$$R^{\uparrow/\downarrow}(x) = \int_0^{eV_{set}} dE \ \rho^{\uparrow/\downarrow}(x,E) \quad , \qquad (S17)$$

$$\widetilde{w}_{N/P} = \frac{w_{N/P,\uparrow}}{w_{N/P,\downarrow}}. \qquad (S18)$$

The experimentally relevant condition of $\Delta \ll eV_{set} \ll v, M$ allows us to approximate $R^{\uparrow/\downarrow}(x) \simeq eV_{set} \rho_N^{\uparrow/\downarrow}(x)$ and the spin contrast from the conductance measured with two tip polarizations ($N$ and $P$) is given by

$$\delta G(E) = G_N(E) - G_P(E) \simeq \frac{(\widetilde{w}_N - \widetilde{w}_P)\left(\tilde{\rho}^{\uparrow}(E) - \tilde{\rho}^{\downarrow}(E)\right)}{\left(\widetilde{w}_N \frac{\rho_N^{\uparrow}(E)}{\rho_N^{\downarrow}(E)} + 1\right)\left(\widetilde{w}_P + \frac{\rho_N^{\downarrow}(E)}{\rho_N^{\uparrow}(E)}\right)} \quad . \qquad (S19)$$

By applying the condition of $\tilde{\rho}^{\uparrow}(E) = \tilde{\rho}^{\downarrow}(-E)$ shown earlier, we find that $\delta G$ satisfies the energy anti-symmetric condition $\delta G(E) = -\delta G(-E)$ and, especially, $\delta G$ as well as $P(E)$ (defined in the main text) vanish at zero energy. This important feature of the in-gap state allows us to distinguish trivial zero energy Shiba states from MZMs.

### Section 7. Numerical simulation of magnetic impurity chain

The details of the spin properties of the in-gap states and the MZM induced by a chain of magnetic impurity have been examined theoretically in a parallel study (*32*). Here we briefly describe the numerical calculation that was performed in this work. We reproduce it here to compare with experimental measurements. These calculations allow us to simulate the properties of the system more quantitatively than in analytical models described in ref (*32*). For these simulations, we consider a chain composed of 60 magnetic impurity sites embedded in a two-dimensional infinite superconductor. We only consider nearest neighboring couplings between the impurity sites and the superconducting sites. The discretized version of the Hamiltonian with generic dispersion (shown in Fig. 5C) is used in this numerical study.

$$\widehat{H} = \widehat{H}_s + \widehat{H}_d + \widehat{H}_T \, , \qquad (S20)$$

$$\widehat{H}_s = \sum_{k_x, k_y} \left( \boldsymbol{c}_{\boldsymbol{k}}^\dagger \quad \overline{\boldsymbol{c}}_{\boldsymbol{k}}^\dagger \right) \, H_s(\boldsymbol{k}) \begin{pmatrix} \boldsymbol{c_k} \\ \overline{\boldsymbol{c}}_{\boldsymbol{k}} \end{pmatrix}, \qquad (S21)$$

$$\widehat{H}_d = \sum_{k_x} \left( \boldsymbol{d}_{k_x}^\dagger \quad \overline{\boldsymbol{d}}^\dagger{}_{k_x} \right) \, H_d(k_x) \begin{pmatrix} \boldsymbol{d}_{k_x} \\ \overline{\boldsymbol{d}}_{k_x} \end{pmatrix}, \qquad (S22)$$

$$\widehat{H}_T = \sum_{\langle r_s, r_d \rangle} V \big( \boldsymbol{c}_{\boldsymbol{r}_s}^\dagger \boldsymbol{d}_{\boldsymbol{r}_d} - \overline{\boldsymbol{c}}_{\boldsymbol{r}_s}^\dagger \overline{\boldsymbol{d}}_{\boldsymbol{r}_d} \big) + h.c. \, , \qquad (S23)$$

where

$$H_s(k) = \left[ 2t_s \big( 2 - \cos k_x - \cos k_y \big) - \mu_s \, + t_{SO} \big( \sin k_x \sigma_y - \sin k_y \sigma_x \big) \right] \otimes \tau_z + \Delta \tau_x \, , (S24)$$

$$H_d(k_x) = \, M \sigma_z + (2t_d \cos k_x - \mu) \tau_z \, , \qquad (S25)$$

where $\boldsymbol{r}_s$ and $\boldsymbol{r}_d$ are the positions of the superconductor and the magnetic impurity sites, respectively. The Hamiltonian defines a quantum Shiba band, where the d-orbital chain with band-structure $H_d(k_x)$ - which includes the spin splitting of the bands in Fe – of order 2 eV, calculated by DFT – couples with the underlying superconductor through a term $\widehat{H}_T$ which hops between the Fe and Pb orbitals. This one orbital Fe and Pb model is not the most realistic model of the triple Fe chain coupled to the Pb superconductor, but has the advantage of being transparent. More detailed calculations, which take into account the triple chain nature of the Fe atomic chain as well as the many orbitals at the Fermi level in Fe and Pb have been performed, and are under submission.

The spectral functions and the spin DOS are numerically calculated from the Green's function of the d-orbital electrons. The spin resolved DOS at the chain end, shown in Fig. 5D, indicates that the MZM mainly comes from the minority spin component, with a small contribution from the majority spin component due to the combined effect of broadening and spin-orbit coupling (15). Shiba states also develop a pair of peaks as in the single impurity case but broader in energy. The normal spin DOS, defined as $\rho_N^{\uparrow/\downarrow}$, is the average spin polarization of the Shiba states over the full k space due to their extended nature. In Fig. 4C & 4D, the simulated conductance $G_{N/P}(x, E)$ and the corresponding polarization $P(x, E)$ are shown for the end and the middle of the chain (where $\widetilde{w}_N = 1.5$ and $\widetilde{w}_P = 0.67$ are used in this simulation).

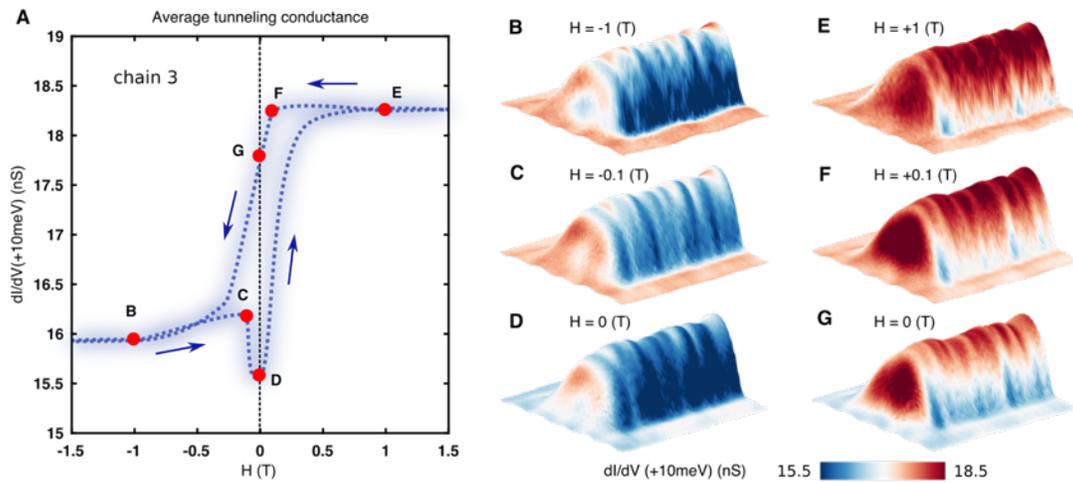

**Fig. S1.**
Magnetic hysteresis of a spin-polarized probe tip. (A) Average tunneling conductance of chain 3 as a function of out-of-plane external field. Experimentally obtained data are marked as red dots. Dashed lines are guide by eye. Similar data with denser data points can be found in the previous work (9). (B to G) Topography of the chain false color mapped by the conductance at 10meV at different magnetic fields ($I_{set}$=750 pA, $V_{set}$=10 mV, $V_{mod}$=1 mV).

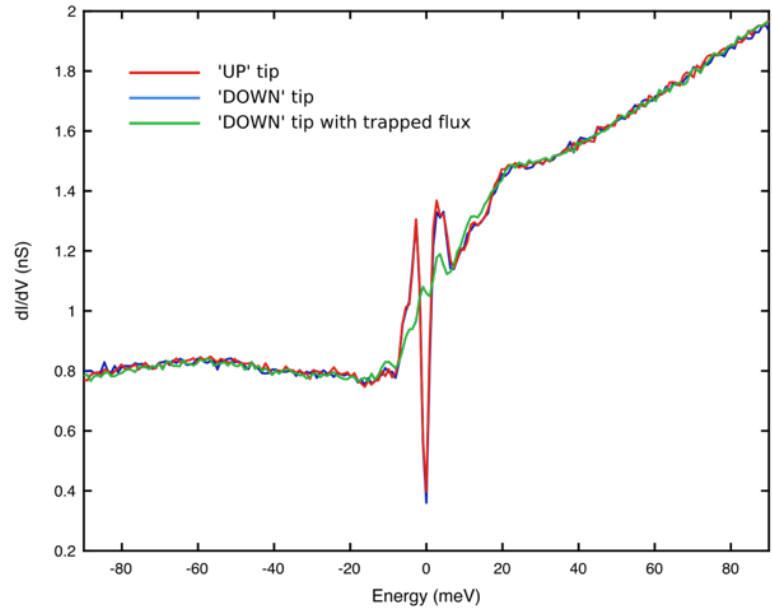

**Fig. S2**

Tunneling spectra measured on Pb(110) with and without trapped magnetic flux. Spectra measured on Pb(110) with 'UP' (red curve) and 'DOWN' (blue and green curves) polarized tip ($V_{set}$=90 mV, $I_{set}$=1 nA, $V_{mod}$=1 mV). The green curve shows suppressed superconducting gap originated from trapped magnetic flux. The red and blue spectra do not show a full superconducting gap due to the strong AC modulation ($V_{mod}$=1 mV).

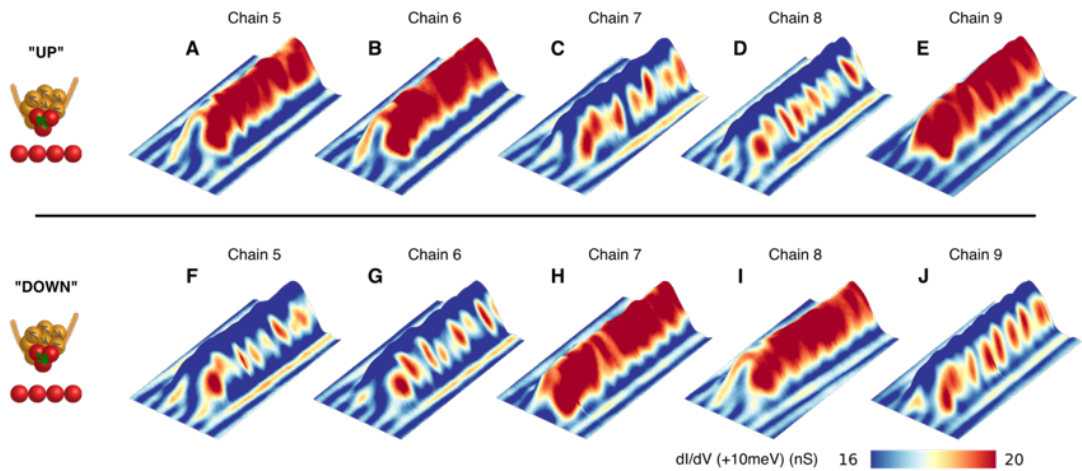

**Fig. S3**

Magnetization of Fe chains measured with 'UP' and 'DOWN' polarized tips. (A to E) Tunneling conductance (10 meV) on Fe chains (from chain 5 to chain 9 accordingly) measured with 'UP' polarized tip ($I_{set}$=750 pA, $V_{set}$=10 mV, $V_{mod}$=1 mV). (F to J) Tunneling conductance (10 meV) on Fe chains (from chain 5 to chain 9 accordingly) measured with 'DOWN' polarized tip ($I_{set}$=750 pA, $V_{set}$=10 mV, $V_{mod}$=1 mV).

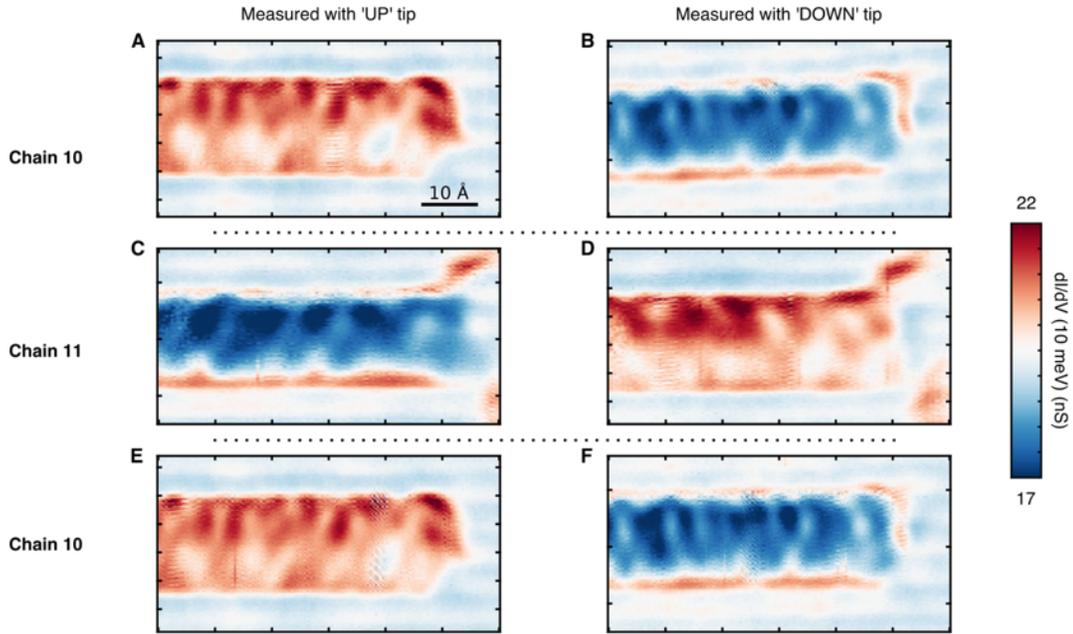

**Fig. S4**

Series of spin polarized measurements in time order. (A and B) Tunneling conductance maps (10 mV) measured on chain 10 with 'UP' (A) and 'DOWN' (B) polarized tip ($V_{set}$=10 mV, $I_{set}$=750 pA, $V_{mod}$=1 mV). (C and D) Tunneling conductance maps (10 mV) measured on chain 11 with 'UP' (C) and 'DOWN' (D) polarized tip ($V_{set}$=10 mV, $I_{set}$=750 pA, $V_{mod}$=1 mV). The polarization of the chain 11 is opposite to that of chain 10. (E and F) Tunneling conductance maps (10 mV) measured on chain 10 with 'UP' (E) and 'DOWN' (F) polarized tip which show the same contrast as (A) and (B) respectively ($V_{set}$=10 mV, $I_{set}$=750 pA, $V_{mod}$=1 mV).

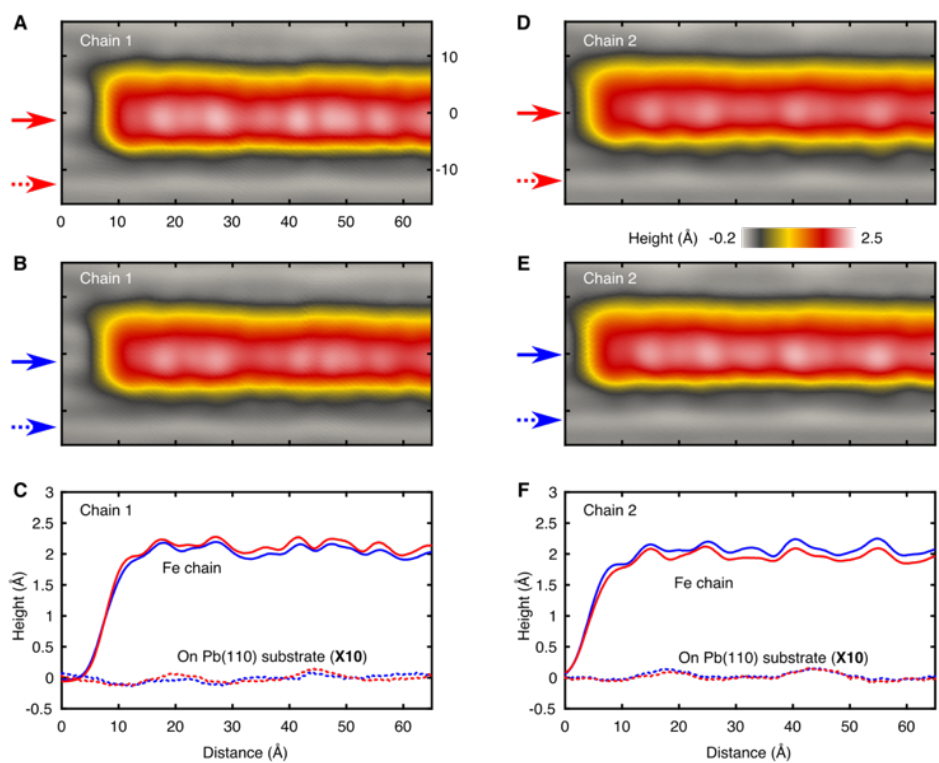

**Fig. S5**

Topographies of chain 1 and 2 measured with 'UP' and 'DOWN' polarized tips. (A and B) Topographic images of chain 1 measured with 'UP' (A) and 'DOWN' (B) polarized tips ($V_{set}$=-5 mV, $I_{set}$=500 pA). (C) The height profiles taken on Fe chain (solid lines) and on Pb(110) substrate (dashed lines) wiht 'UP' (red) and 'DOWN' (blue) polarized tip. The height profile for Pb(110) is amplified by 10 times. (D and E) Topographic images of chain 2 measured with 'UP' (D) and 'DOWN' (E) polarized tips ($V_{set}$=-5 mV, $I_{set}$=500 pA). (F) The height profiles taken on Fe chain (solid lines) and on Pb(110) substrate (dashed lines) wiht 'UP' (red) and 'DOWN' (blue) polarized tip. The height profile for Pb(110) is amplified by 10 times.

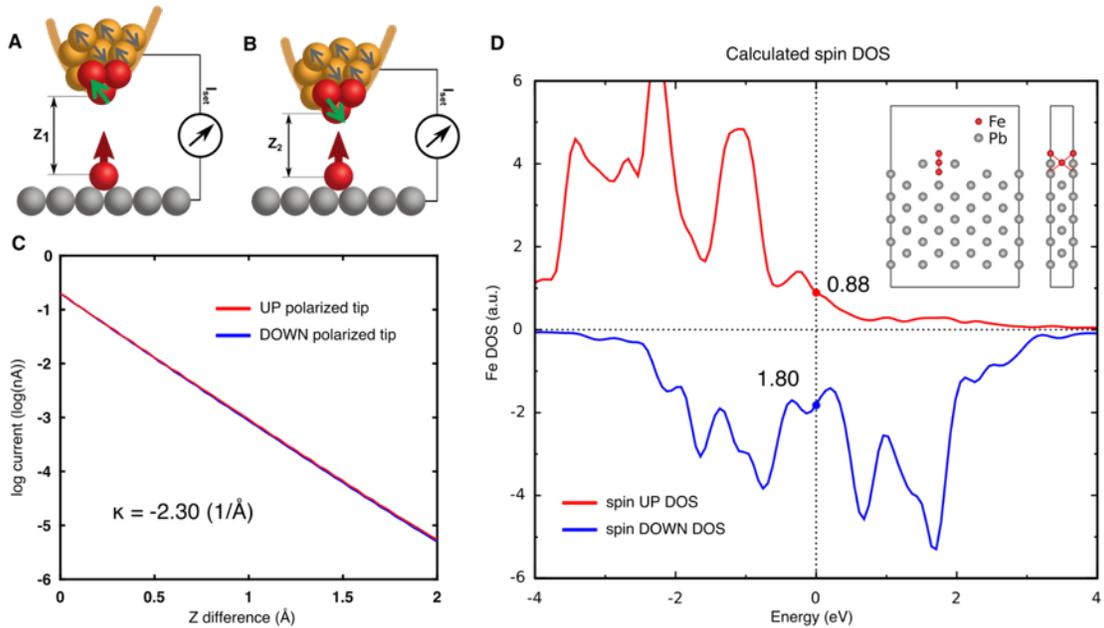

**Fig. S6**

Normal state spin-polarization and spin-resolved DFT calculation on Fe chain. (A and B) Schematic of spin-polarized measurement with parallel (A) and anti-parallel (B) magnetic configurations between tip and sample. The distance between tip and sample is adjusted by the STM feedback loop. (C) dI/dZ measurement on Fe chain. Both up and down polarized tips show the same slope of -2.30 (1/Å). (D) Spin resolved DFT calculation of Fe chain on Pb substrate. The red (blue) curve stands for the spin up (down) DOS. Inset shows the calculated geometry where red and grey dots represent the position of Fe and Pb atoms, respectively.

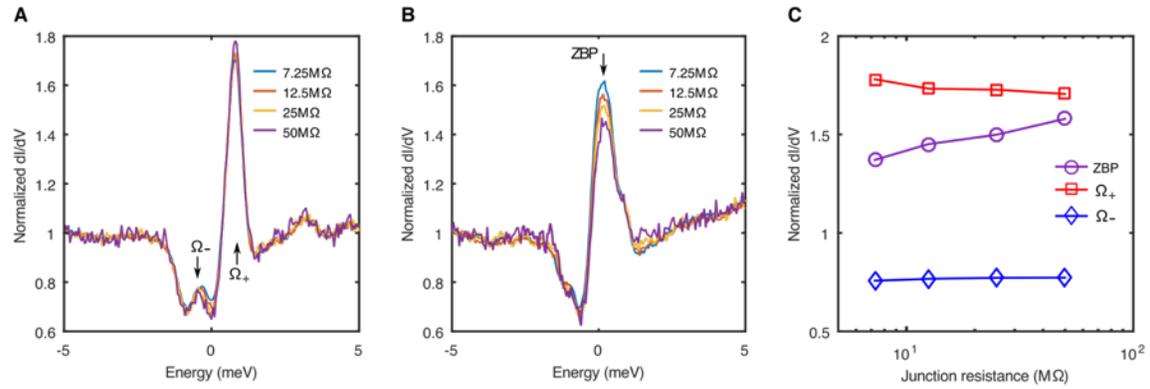

**Fig. S7**

Junction-resistance dependent conductance. (A and B) Normalized conductance spectra in the middle of chain (A) and at the end of chain (B) with different junction resistances ($V_{set}$=-5 mV, $V_{mod}$=40 μV). $\Omega_+$ and $\Omega_-$ denote the van-hove singularity energy of the Shiba band. (C) Normalized conductance of ZBP (purple circles), $\Omega_+$ (red squares), and $\Omega_-$ (blue diamond) as a function of junction resistance.

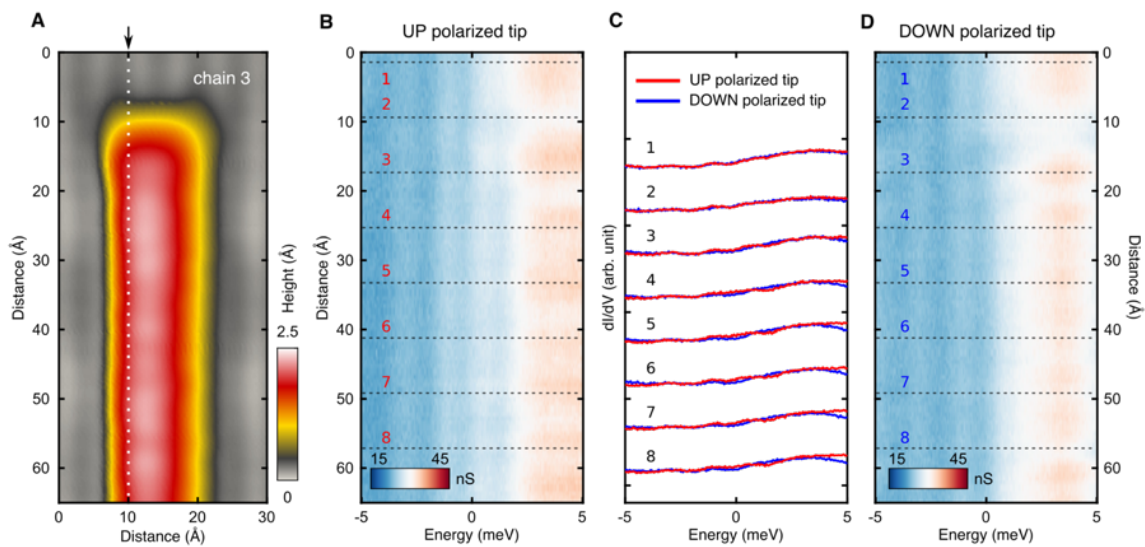

**Fig. S8**

Spatially resolved spin-polarized state on a Fe atomic chain at 100mT. (A) Topographic image of chain 3 taken at 100 mT (out-of-plane field, $V_{set}$=-10 mV, $I_{set}$=750 pA). (B and D) Spatial variation of the spectra taken along the white dashed line shown in (A) with UP (B) and DOWN (D) polarized tips ($V_{set}$=-5 mV, $I_{set}$=500 pA, $V_{mod}$=40 µV). (C) Individual spectra from (B) (red curves) and (D) (blue curves) at the labeled locations.

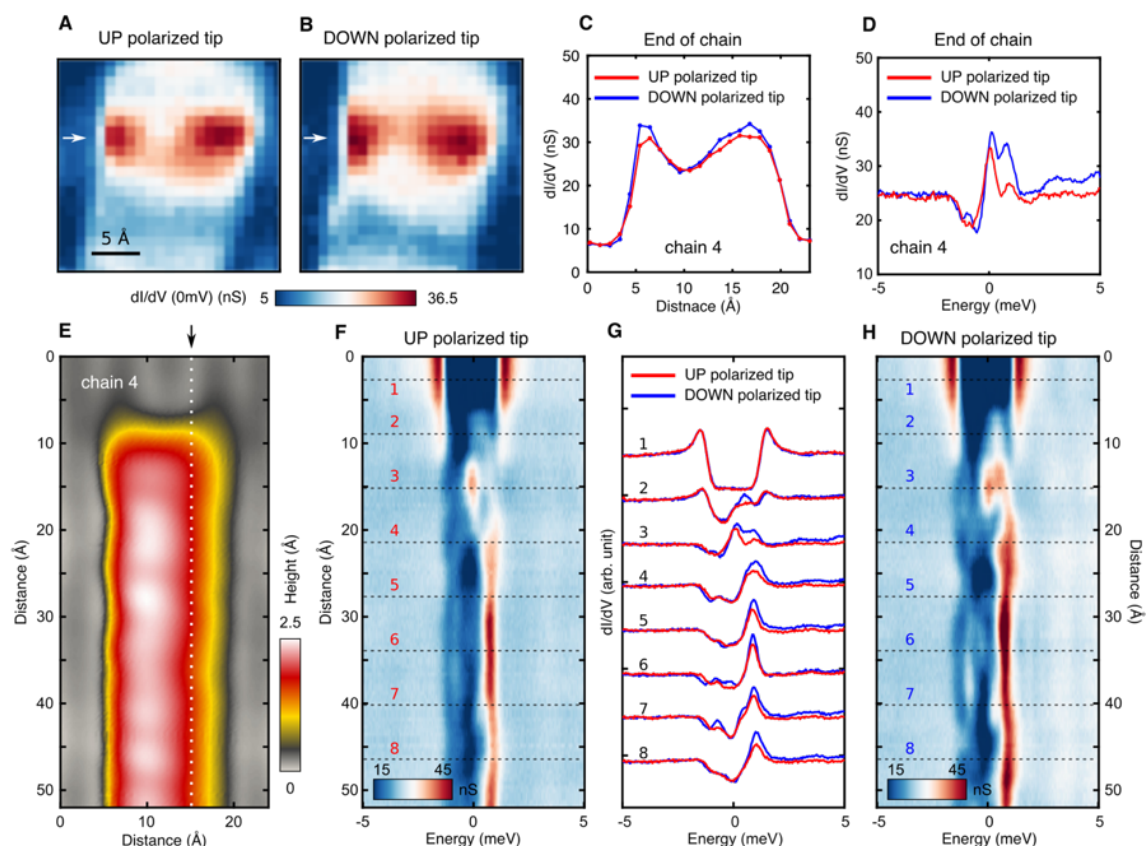

**Fig. S9**

Spin-polarized state on a Fe atomic chain with opposite ferromagnetic orientation. (A and B) Zero energy conductance map near the chain end taken with UP (A) and DOWN (B) polarized tips ($V_{set}$=-5 mV, $I_{set}$=500 pA , $V_{mod}$=40 μV). (C) Line cuts taken from (A) and (B) across the double-eye feature measured with UP (red curves) and DOWN (blue curves) polarized tips. (D) Spectra at the end of chain ($V_{set}$=-5 mV, $I_{set}$=500 pA, $V_{mod}$=40 μV). Down polarized tip shows higher conductance than up polarized tip. Notice the reversal of spin-polarization compared to the chain 3 in the main figures. (E) Topographic image of chain 4 ($V_{set}$=-10 mV, $I_{set}$=750 pA). (F and H) Spatial variation of the spectra taken along the white dashed line shown in (E) with UP (F) and DOWN (H) polarized tip ($V_{set}$=-5 mV, $I_{set}$=500 pA, $V_{mod}$=40 μV). (G) Individual spectra from (F) (red curves) and (H) (blue curves) at the labeled locations.

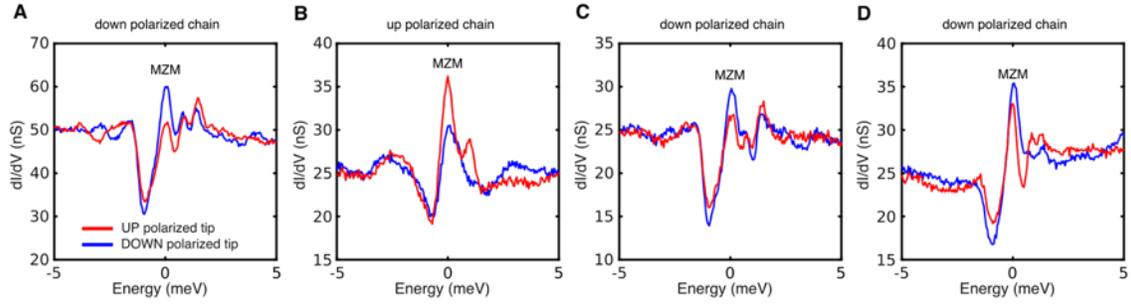

**Fig. S10**

More examples of Spin-polarized in-gap states at the end of chains. (A to D) Conductance spectra measured at the edge of different chains with different tips ((A) $V_{set}$=5 mV, $I_{set}$=1 nA, $V_{mod}$=40 μV. (B) $V_{set}$=5 mV, $I_{set}$=500 pA, $V_{mod}$=40 μV. (C) $V_{set}$=5 mV, $I_{set}$=500 pA, $V_{mod}$=40 μV. (D) $V_{set}$=-5 mV, $I_{set}$=500 pA, $V_{mod}$=40 μV). The spectra in (B) is measured on 'UP' polarized chain, while the other spectra (A, C, and D) are taken on 'DOWN' polarized chains.